\documentclass[12pt,preprint]{aastex}
\usepackage{graphicx}
\usepackage{amssymb}
\usepackage{lscape}
\usepackage{amsmath}

\shorttitle{The Highly Collimated Radio Jet in the HH 80-81 Protostellar Complex}
 \shortauthors{Rodriguez-Kamenetzky et al.}

\begin{document}

\title{The Highly Collimated Radio Jet of HH 80-81: Structure and Non-Thermal Emission}

 \author{Adriana Rodr\'{\i}guez-Kamenetzky\altaffilmark{1,2}, Carlos Carrasco-Gonz\'alez\altaffilmark{2}, Anabella Araudo\altaffilmark{3}, Gustavo E. Romero\altaffilmark{4}, Jos\'e M. Torrelles\altaffilmark{5 \dagger}, Luis F. Rodr\'{\i}guez\altaffilmark{2}, Guillem Anglada\altaffilmark{6}, Josep Mart\'{\i}\altaffilmark{7}, Manel Perucho\altaffilmark{8}, Carlos Valotto\altaffilmark{1}}

  \altaffiltext{1}{Instituto de Astronom\'ia Te\'orica y Experimental, (IATE-UNC), X5000BGR C\'ordoba, Argentina} 
  \altaffiltext{2}{Instituto de Radioastronom\'ia y Astrof\'isica (IRyA-UNAM), 58089 Morelia, M\'exico}
  \altaffiltext{3}{Astronomical Institute of the Academy of Sciences, Prague, Czech Republic}
  \altaffiltext{4}{Instituto de Argentino de Radioastronom\'ia (IAR-CONICET \& CIC), C.C. 5, 1894, Villa Elisa, Buenos Aires, Argentina}
  \altaffiltext{5}{Institut de Ci\`{e}ncies de l'Espai (CSIC-IEEC) and Institut de Ci\`{e}ncies del Cosmos (UB-IEEC), Carrer de Can Magrans S/N, 08193 Cerdanyola del Vall\`{e}s, Barcelona, Spain}
  \altaffiltext{6}{Instituto de Astrof\'{\i}sica de Andaluc\'{\i}a, CSIC, Camino Bajo de Hu\'etor 50, E-18008 Granada, Spain}
  \altaffiltext{7}{Dept. de F\'{\i}sica, EPS de Ja\'en, Universidad de Ja\'en, Campus Las Lagunillas s/n, A3-402, 23071 Ja\'en, Spain}
  \altaffiltext{8}{Dept. d'Astronomia i Astrof\'{\i}sica, Universitat de Val\`encia, C/ Dr. Moliner 50, 46100, Burjassot (Val\`encia), Spain and Observatori Astron\`omic, Universitat de Val\`encia, C/ Catedr\`atic Jos\'e Beltr\'an 2, 46980, Paterna (Val\`encia), Spain}
  \altaffiltext{$\dagger$}{The ICC (UB) is a CSIC-Associated Unit through the ICE}

\begin{abstract}

Radio emission from protostellar jets is usually dominated by free-free emission from thermal electrons. However, in some cases, it has been proposed that non-thermal emission could also be present. This additional contribution from non-thermal emission has been inferred through negative spectral indices at centimeter wavelengths in some regions of the radio jets. In the case of HH 80-81, one of the most powerful protostellar jets known, linearly polarized emission has also been detected, revealing that the non-thermal emission is of synchrotron nature from a population of relativistic particles in the jet. This result implies that an acceleration mechanism should be taking place in some parts of the jet. Here, we present new high sensitivity and high angular resolution radio observations at several wavelengths (in the 3-20 cm range) of the HH80-81 radio jet. These new observations represent an improvement in sensitivity and angular resolution by a factor of $\sim$10 with respect to previous observations. This allows us to resolve the morphology of the radio jet, and to study the different emission mechanisms involved through spectral index maps. We conclude that synchrotron emission in this jet arises from an extended component detected at low frequencies and from the termination points of the jet, where strong shocks against the ambient medium can produce efficient particle acceleration.
\end{abstract}


\section{Introduction}

 Jets from young stellar objects (YSOs) are frequently associated with weak centimeter emission (Anglada et al. 1992; Anglada 1995; Rodr\'iguez \& Reipurth 1998). In the best studied cases, these sources are resolved angularly at the sub-arcsec scale and found to be elongated in the direction of the large-scale tracers of the jet (e.g. Rodr\'{\i}guez et al. 1990; Rodr\'{\i}guez 1995, 1996; Anglada 1996), indicating that they trace the region, very close to the exciting star, where the outflow phenomenon originates. The centimeter flux density ($S_{\nu}$) usually rises slowly with frequency ($\nu$), i.e., it exhibits a positive spectral index $\alpha$ (defined as $S_\nu \propto \nu^\alpha$). Given these characteristics, the radio emission is usually interpreted as free-free emission from ionized particles in the jet, and the radio sources are usually referred to as thermal radio jets.
 
In past decades, radio emission with negative spectral indices at centimeter wavelengths has been reported in several YSO jets, such as the triple radio source in Serpens (Rodr\'iguez et al. 1989, Rodr\'iguez-Kamenetzky et al. 2016), HH 80-81 (Mart\'{\i} et al. 1993), IRAS~16547$-$4247 (Garay et al. 1996; Rodr\'iguez et al. 2005), W3(H$_2$O) (Wilner et al. 1999), L778-VLA6 (Girart et al. 2002), DG Tau (Ainsworth et al. 2014), NGC6334I-CM2 (Brogan et al. 2016), and OMC-2 FIR 3 (Osorio et al. 2017) suggesting a non-thermal origin for the emission. In a recent radio survey of the southern hemisphere, Purser et al. (2017) found at least 10 new candidates to present non-thermal emission. Non-thermal emission from protostellar jets has been interpreted as synchrotron emission from a small population of relativistic particles, that might be accelerated in strong interactions of the jet with the ambient medium. Since emission from YSO jets has usually been believed to be only of thermal origin, the presence of a non-thermal component contitutes a striking discovery. In the situation of strong shocks of the jet with the ambient medium, particles could gain energy by diffusing back and forth across a shock front (e.g., Krymskii 1977, Axford et al. 1977, Bell 1978a,b, Blandford \& Ostriker 1978) allowing thermal and supra-thermal particles to reach relativistic velocities, through the mechanism known as Diffusive Shock Acceleration (DSA). This process is well known to work in relativistic jets (e.g., Blandford et al. 1982), supernova remnants (e.g., Castro \& Slane 2010), nova ejecta (e.g., Kantharia et al. 2014), and colliding wind binaries (e.g., De Becker 2007), where the velocities involved in shocks are of at least several thousand kilometers per second. Nevertheless, motivated by the detection of synchrotron emission in (few) YSO jets, a number of authors have also studied DSA as a possible mechanism to accelerate particles in these jets moving with typical velocities of only a few hundreds kilometers per second (Crusius-Watzel 1990, Araudo et al. 2007, Bosch-Ramon et al. 2010, Padovani et al. 2015, 2016).

The first conclusive evidence for the presence of synchrotron emission in jets from YSOs was given by Carrasco-Gonz\'alez et al. (2010), with the detection of linearly polarized radiation in one of the most powerful protostellar jets known, HH~80-81. This discovery has very important implications: a) magnetic fields in these objects can be studied with techniques usually applied to relativistic jets; b) high-energy phenomena should be present, and hence, jets from young stars could be a new ground to study DSA in low-velocity and high-density plasmas, i.e. in a regime that has not been explored in deep yet. More recently, Rodr\'{\i}guez-Kamenetzky et al. (2016) reported a multifrequency study of the triple radio source in Serpens, where very sensitive radio observations revealed a clear difference in the emission nature of the jet (non-thermal) and its exciting source (thermal). These authors also found that particle acceleration in that object might occurs via DSA in strong shocks against the ambient medium, only requiring a jet slightly faster than usual (jet velocities larger than 600~km~s$^{-1}$, while typical velocities in YSO jets are in the range 100-300~km~s$^{-1}$). All these results suggest that jets from YSOs, at least under certain conditions, are able to accelerate particles up to relativistic energies and generate synchrotron emission.

In this work we focus our attention on the powerful HH 80-81 protostellar jet, located at the edge of the L291 cloud in Sagittarius at a distance of 1.7 kpc (Rodr\'iguez et al. 1980). Since the early 1980s, the detection of a compact radio continuum source and H$_2$O maser emission (Rodr\'iguez et al. 1980) called the attention to this region as an important site of recent star formation, which has been the subject of numerous studies. In 1985 the IRAS survey detected the bright IR source IRAS 18162-2048, associated with the compact radio continuum emission. Because of its high luminosity (1.7 x 10$^4$ L$_{\odot}$), it is believed to be the case of a massive protostar or protostellar cluster (Aspin \& Geballe 1992). This region contains the Herbig-Haro (HH) objects 80 and 81 (Reipurth \& Graham 1988), among the brightest HH objects known. Interferometric observations at 6 and 3.6~cm revealed a highly collimated bipolar radio jet emanating from the protostar (Mart\'i, Rodr\'iguez \& Reipurth 1993), with velocities of the order of 1000 km~s$^{-1}$ (Mart\'i et al. 1995). A disk with a radius of a few hundreds au was detected around the central star (Carrasco-Gonz\'alez et al. 2012, Fern\'andez-L\'opez et al. 2011). In addition to these underlying features, the detection of linearly polarized emission in the radio continuum (Carrasco-Gonz\'alez et al. 2010) revealed its synchrotron origin, implying the presence of relativistic particles in a non-relativistic medium, and therefore, the action of an acceleration mechanism.

Here we present new data at radio continuum wavelengths of the HH 80-81 inner jet (i.e., within 45 arcsec from its exciting source) observed with the Jansky Very Large Array radio interferometer, and discuss the results in the context of particle acceleration and synchrotron radiation. Our new observations represent an improvement in sensitivity and angular resolution by a factor of $\sim$10 with respect to the previous observations by Carrasco-Gonz\'alez et al. (2010), allowing to better resolve the morphology of the HH 80-81 radio jet, and to study the different emission mechanisms involved through spectral index maps.

 \section{Observations}\label{obs}

Observations of the HH 80-81 jet were made with the Karl G. Jansky Very Large  Array (VLA) of the National Radio Astronomy Observatory (NRAO)\footnote{The NRAO is a facility of the National Science Foundation operated under cooperative agreement by Associated Universities, Inc.}. We observed the continuum emission in the L, S, and C bands in B configuration during June 12 and 16, 2012 (project code: 12A-240). The covered frequency ranges were 1-2 GHz (L band), 2-4 GHz (S band), and 4-6 GHz (C band); however, due to strong RFIs, the final usable bandwidth at L and S bands were 1.3-1.7 GHz and 2.4-3.6 GHz, respectively.
Each band is divided in 1024 channels of 1 MHz for L band and 2 MHz for S and C bands. Bandpass and flux calibration were made by observing 3C286. Complex gain calibration was achieved by observation of J1911-2006. We also performed polarization calibration by using 3C286 as the polarization angle calibrator and the source J1824+1044 as the leakage calibrator, that was observed several times at different parallactic angles. We found an upper limit for the polarization degree of 30$\%$ in L, S, and even in C band, where Carrasco-Gonz\'alez et al. (2010) detected polarized emission over 10$\%$. This is because, despite the improved rms in our C band images ($\sim$ 5~$\mu$~Jy/beam, half the rms in Carrasco-Gonz\'alez et al. 2010), the flux density per beam is significantly lower due to the higher angular resolution reached in our observations ($\sim$ 1.5$''$, in contrast with the angular resolution of the Carrasco-Gonz\'alez et al. 2010 observations, $\sim$10$''$). Therefore, more sensitive observations are needed in order to detect polarized emission and study magnetic fields in images with angular resolution of a few arcsec where the jet structure can be resolved.

The phase center of our observations was $\alpha$(J2000)=18$^h$19$^m$12.1$^s$, $\delta$(J2000)=$-$20$^{\circ}$47$'$30.9$''$.
Calibration of the data was undertaken with the data reduction package CASA (Common Astronomy Software Applications\footnote{https://science.nrao.edu/facilities/vla/data-processing}; version 4.5.0) following standard VLA procedures. Cleaned images were made using the task \emph{clean} of CASA, using multifrequency synthesis and multiscale cleaning (Rau \& Cornwell 2011). For these data, we made images selecting different bandwidths: 0.4 GHz (L band, $\lambda_{L}=18.4$~cm), 1.2 GHz (S band, $\lambda_{S}=10$~cm), and 2 GHz (C band, $\lambda_{C}=5.5$~cm), and also images combining data from different bands. We used different values of the parameter \emph{robust} of \emph{clean} (Briggs 1995), ranging from $-$2 (uniform weighting) to $+$2 (natural weighting). Image parameters are summarized in Table \ref{tbl-images}.

\section{Results and discussion}\label{resultados}

\begin{figure}
  \epsscale{1.0}
  \plotone{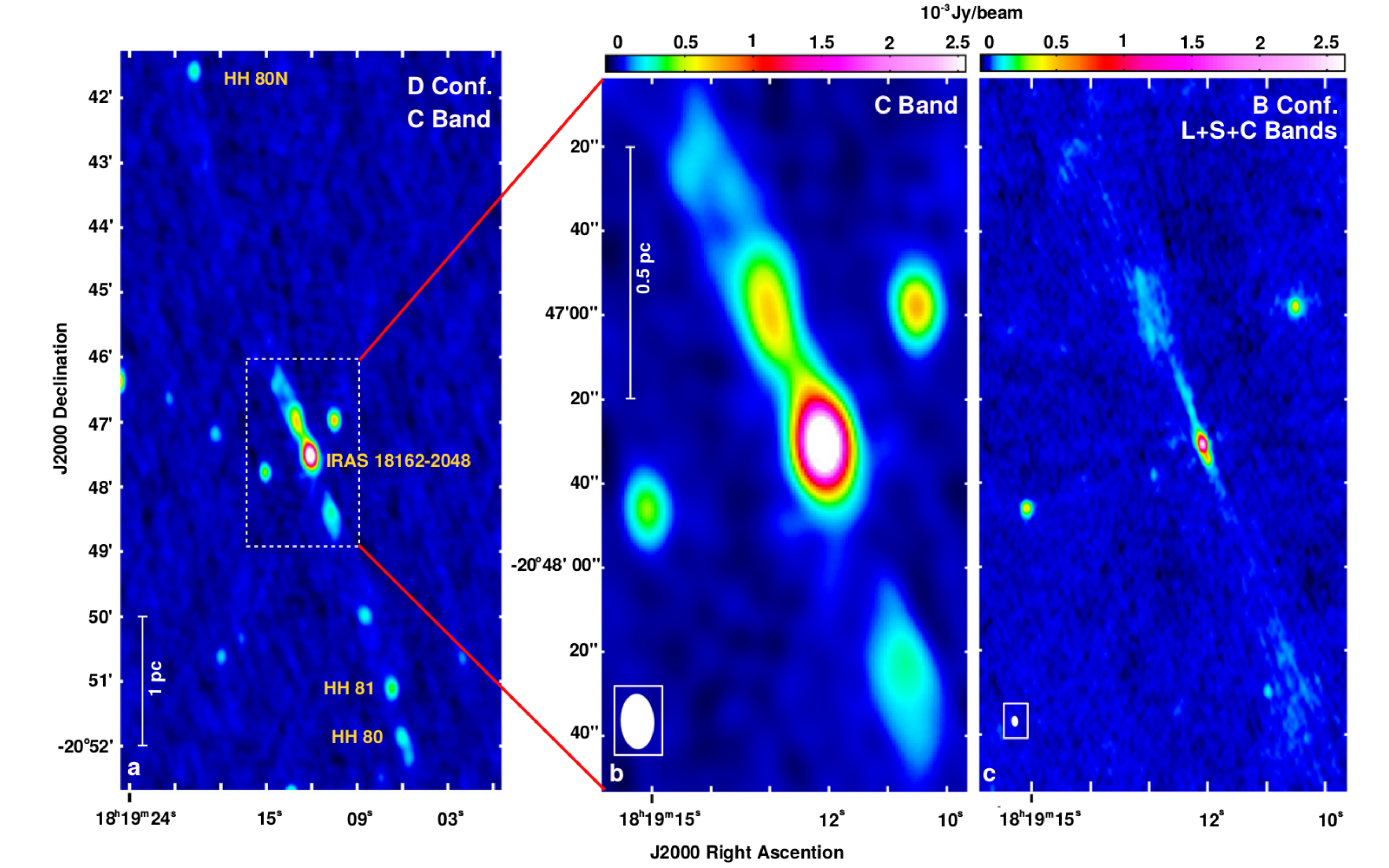}
  \caption{\footnotesize{Radio continuum images of the HH 80-81 jet region. (a) and (b) panels correspond to the C band image, obtained with the JVLA in D configuration by Carrasco-Gonz\'alez et al. (2010), while (c) panel corresponds to our widest bandwidth (L+S+C bands) image taken in B configuration with an angular resolution of $\sim$2.37$''$$\times$1.46$''$ (PA = 4$^{\circ}$) (this paper). The whole extension of the HH 80-81 jet is seen in panel (a), where the HH objects and the source associated with the central protostar (IRAS 18162-2048) are labeled. The structure of the internal region of the jet (panel b) appears poorly resolved in the VLA-D configuration image, but our new high angular resolution image (panel c) reveals a highly collimated jet finishing in two lobe-like structures (see also Fig. \ref{Fig2}).}}
  \label{Fig1} 
  \end{figure}

 In order to show the region we studied, we introduce the HH 80-81 radio jet in Figure \ref{Fig1}. In the left panel we show the C band image, previously reported by Carrasco-Gonz\'alez et al. (2010), showing the whole extension of the radio jet ($\sim$5.5~pc). The central panel shows the inner part of the jet with relatively low angular resolution (13$''$ $\times$ 8$''$, PA= 2$^{\circ}$). The same region is shown in our new high sensitivity multi-band image (right panel), with a greatly improved angular resolution ($2.37''\times 1.46''$, PA = $-$176$^{\circ}$). Our new image reveals a highly collimated bipolar jet finishing in two lobe-like structures. In the following, we discuss about the results obtained from these new radio data. Thus, in subsection \ref{structure}, we discuss about the newly revealed structure of the radio jet. In subsection \ref{nature} we use spectral index maps of the radio emission in order to  discuss about the different emission mechanisms present. In this subsection, we also discuss about the relationship between the different emission mechanism and different structures in the radio jet. A more detailed discussion on the nature of the emission in the collimated part of the radio jet is presented in subsection \ref{collimated}. Finally, in subsection \ref{other} we report four new compact radio sources which have been identified for the first time in these new data.

\subsection{Jet structure at centimeter continuum wavelengths} \label{structure}

 In order to study the structure of the radio jet with our new high angular resolution images, we compare images at different wavelengths and different resolutions. Low frequency images are more sensitive to extended structures, while high frequency images tend to filter extended emission and detect only compact emission. Therefore, by observing in different spectral bands, we are able to study different structures in the jet. 

\begin{figure}
  \epsscale{1}
  \plotone{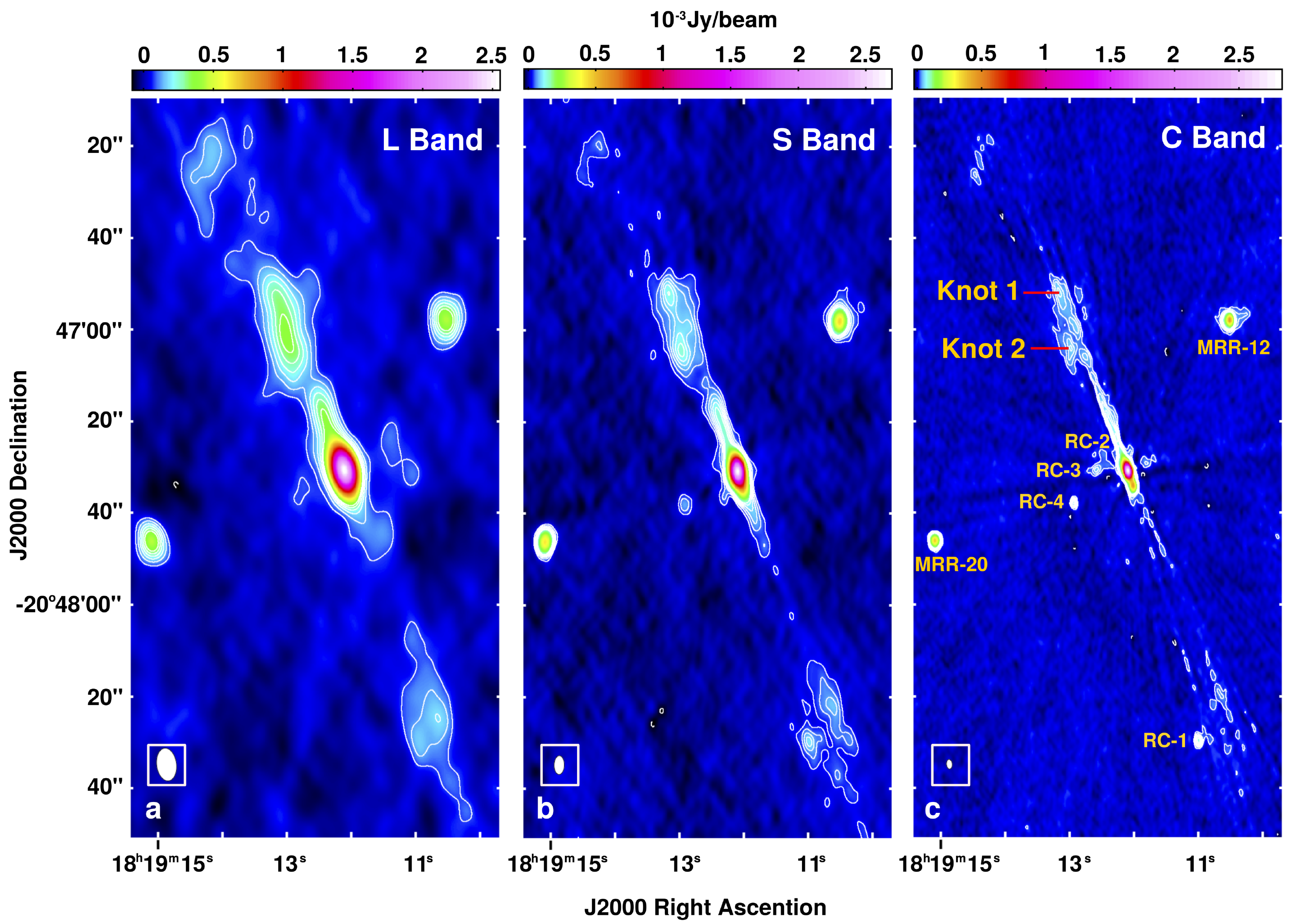}
  \caption{\footnotesize{Radio continuum images at L, S, and C bands (color). Contour levels are 3, 5, 7, 9, 11, and 13 times the rms of each map. (a) L band: rms=~26 $\mu$Jy~beam$^{-1}$; beam = 7$\farcs$17 $\times$ 4$\farcs$30, PA = 7$^\circ$. (b) S band: rms=~10 $\mu$Jy~beam$^{-1}$; beam = 4$\farcs$10 $\times$ 2$\farcs$21, PA = -1$^\circ$. (c) C band: rms=~6 $\mu$Jy~beam$^{-1}$; beam = 2$\farcs$09 $\times$ 1$\farcs$26, PA = 5$^\circ$. We also labeled the two high intensity spots in the lobe-like structure (knot 1 and knot 2), the new sources identified (RC-1, RC-2, RC-3, and RC-4), and two background sources (MRR-12 and MRR-20) reported by Mart\'i, Rodr\'iguez, and Reipurth (1993).
  }}\label{Fig2}
  \end{figure}

In Figure \ref{Fig2} we can see that the structure emanating in opposite directions from the central region is detected with a full extension of $\simeq 1~{\rm pc}$ at all bands (L, S, and C) and angular resolutions. However, a change in the morphology is observed when going from low to higher frequencies: at low frequencies and low angular resolution (L band, Fig. \ref{Fig2}) very extended lobes are observed to the northeast and southwest of the central region, while for increasing frequencies and higher angular resolution (S and C bands, Fig. \ref{Fig2}), a very well collimated jet and several high intensity spots on the lobes are revealed. These changes in morphology seem to be real when comparing the extent of the emission with the beam size of each image. In Figure \ref{Fig2} we labeled the two brightest compact knots in the northern lobe (knot 1 and knot 2) that are interpreted as strong shocks of the jet with the ambient medium (see section \ref{acc}). Both knots are detected with intensities up to nine times the rms noise level.

 
\subsection{Nature of the radio continuum emission} \label{nature}
 
\begin{figure}
   \epsscale{1.0}
   \plotone{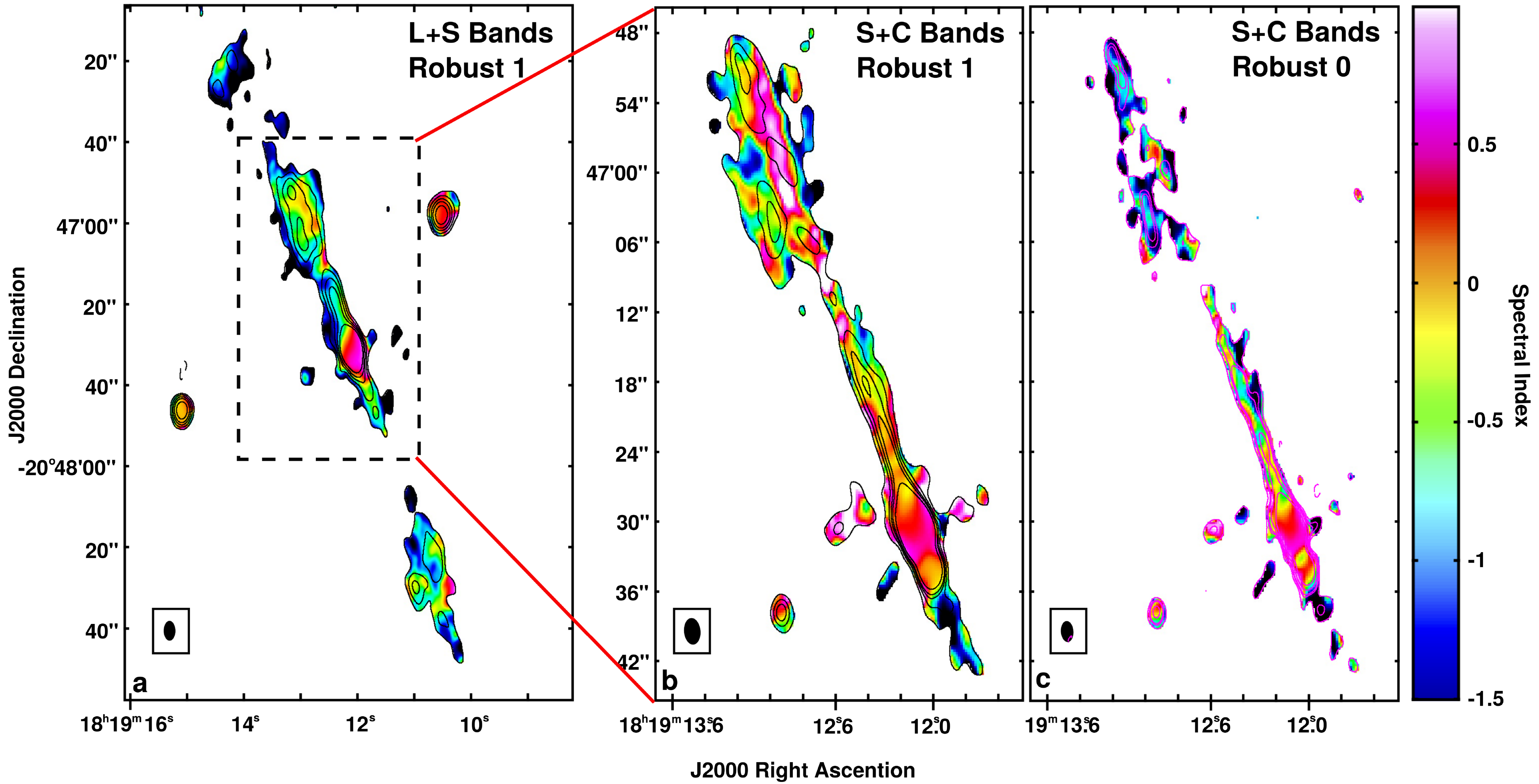}
   \caption{\footnotesize{Superposition of radio continuum images over the spectral index map (color scale) obtained from multifrequency synthesis cleaning. Contours are 3, 5, 8, 12 and 20 times the rms of the continuum image in each panel. (a) Combination of L and S bands: robust=1 weighting, rms=~13 $\mu$Jy~beam$^{-1}$; beam =4$\farcs$81 $\times$ 2$\farcs$73, PA = 0$^\circ$. (b) Combination of S and C bands: robust=1 weighting, rms=~6 $\mu$Jy~beam$^{-1}$; beam=2$\farcs$56 $\times$ 1$\farcs$37, PA = 4$^\circ$. (c) Combination of S and C bands: robust=0 weighting, rms=~6 $\mu$Jy~beam$^{-1}$; beam=1$\farcs$77 $\times$ 1$\farcs$03, PA = 4$^\circ$. The pixels shown are those with S/N $>$ 3.}}
  \label{Fig3}
  \end{figure}

\begin{figure}
  \epsscale{1}
  \plotone{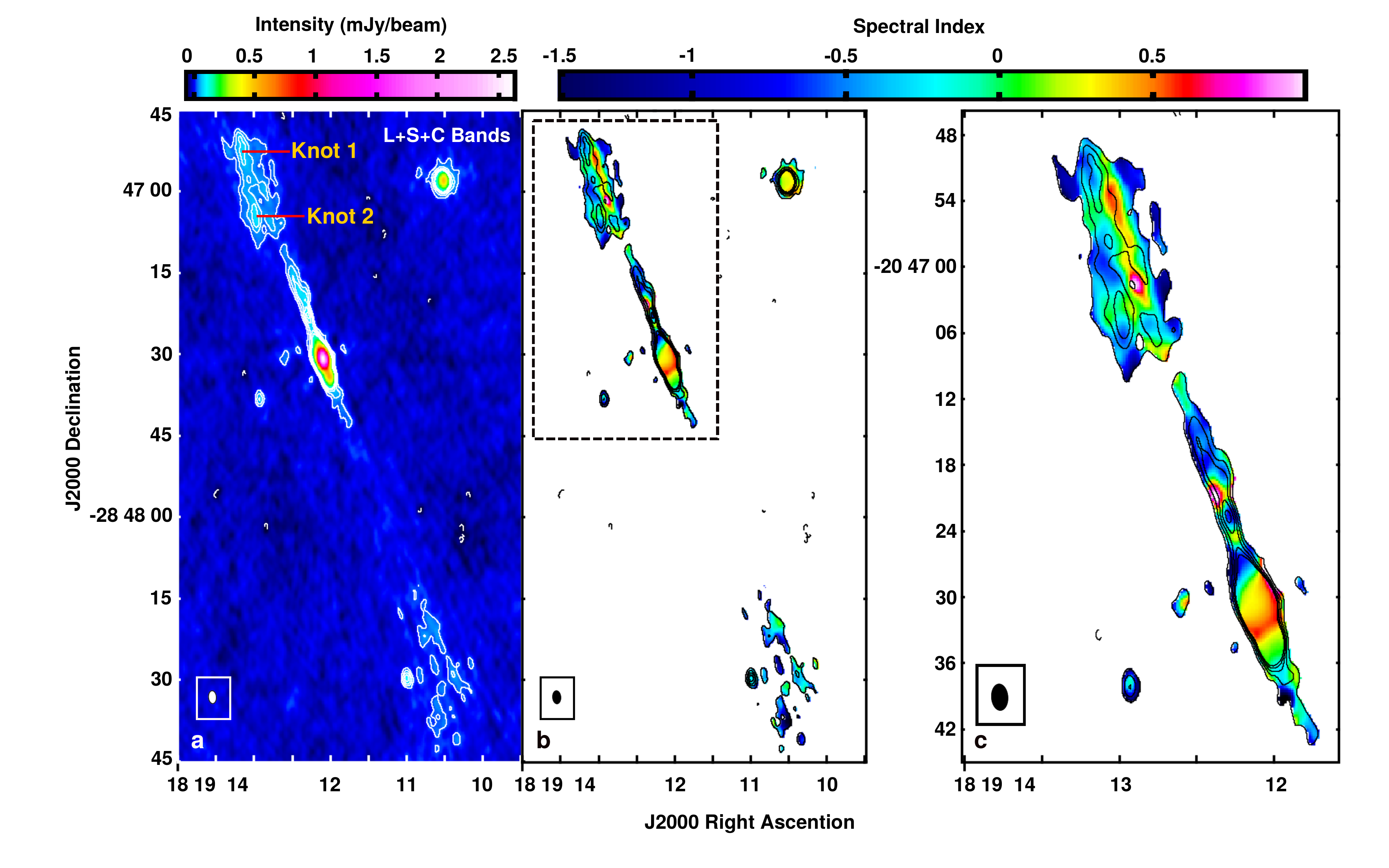}
  \caption{\footnotesize{Radio continuum and spectral index images of the HH 80-81 jet internal region, made by the combination of L, S, and C bands of epoch 2012, obtained from multifrequency synthesis cleaning and robust weighting 1. Beam =2$\farcs$37 $\times$ 1$\farcs$46 (PA=4$^\circ$). (a) Continuum image; contours are 3, 5, 7, 9, and 11 times the rms, 9 $\mu$Jy/beam. Emission peaks (knot 1 and 2) are labeled. (b) Intensity contours of panel (a) over spectral index (color scale). The pixels shown in the spectral index map are those with S/N $>$ 3 in the continuum image. (c) Close-up of the NE internal region of the jet.}}
  \label{Fig4}
  \end{figure}

 In this subsection, we study the different emission mechanisms in the radio jet by comparing spectral index maps made at different frequency ranges and angular resolutions. We combined data from L and S bands to obtain a low frequency map with relatively low angular resolution, which allows us to study the emission mechanism of the extended emission (left panel in Fig. \ref{Fig3}). Similarly, the combination of S and C bands yields a higher angular resolution map which enables the study of the emission mechanisms in the more compact structures (central and right panels in Fig. \ref{Fig3}). Finally, we combined the L, S, and C bands to obtain the highest angular resolution map ($\sim$2$''$), covering a wider frequency range (Fig. \ref{Fig4}).

 {In all the spectral index maps}, we measured a positive spectral index $\alpha=0.33\pm0.02$ for the central region of the jet, suggesting partially optically thick free-free emission, in well agreement with what is expected for a thermal radio jet (Anglada 1996). In contrast, emission from the collimated jet and the northeasten lobe show peculiar variations with frequency and angular resolution.

 Figure \ref{Fig3}a shows that negative spectral indices dominate the extended emission at L+S bands, suggesting a non-thermal nature. An index of $\alpha=-0.5\pm0.4$ is found in high intensity spots of the northeastern lobe (knot 1 and 2), while even more negative values ($\sim-0.8$) are computed in surrounding regions and in the collimated jet ($\sim-0.7$), with typical errors of 0.2. Since synchrotron radiation dominates at low frequencies, it is expected its contribution to be smaller at S+C bands. Indeed, thermal emission seems to be dominant in the S+C bands (Fig. \ref{Fig3}b), where a zone with positive spectral indices can be clearly identified in the northeastern lobe, aligned with the collimated jet. Nevertheless, the high intensity spots, knot 1 and knot 2, present a mixture of flat ($\alpha\simeq$ 0) and negative spectral indices ($\alpha\simeq-0.5$) indicating contribution of both optically thin free-free and non-thermal emission, respectively. When improving the angular resolution in the same frequency range (S+C bands, robust = 0; Fig. \ref{Fig3}c), the emission contribution from the extended component is practically absent and knots 1 and 2 reveal their non-thermal origin ($\alpha=-0.6\pm0.2$). In the same image (Fig. \ref{Fig3}c), a knotty distribution of spectral indices is found in the collimated jet: regions with negative and positive spectral indices alternate.

 In Fig. \ref{Fig4} (L+S+C bands), due to the wide frequency range, the observed emission is expected to be a combination of thermal and non-thermal components. In this image, non-thermal emission dominates in both the northeastern lobe and the jet, with spectral indices $-$1$\leq\alpha\leq$0, and errors of $\sim0.4$. Knots 1 and 2 present spectral indices of $-0.5\pm0.4$. It is worth noting that spectral indices of $-0.6\pm0.1$ are measured very near the thermal source associated with the central massive protostar, in the region where the jet seems to collimate. Close to this area, positive and negative spectral indices alternate, and a high intensity knot with $\alpha=-0.8\pm0.4$ is detected. Errors in the most collimated region of the jet (8 to 23 arcsec from the central region) are of the order of $0.3$. Aligned with the jet, the thermal structure in the northeastern lobe is also present in this range of frequencies.  

In summary, the central part of the jet is consistent with a thermal radio jet, while non-thermal emission seem to be present in different spots in the collimated part of the radio jet, in the high intensity spots which trace the termination shocks of the radio jet, and in an extended emission surrounding the whole radio jet.

\subsection{The collimated jet} \label{collimated}

\begin{figure}
  \epsscale{0.7}
  \plotone{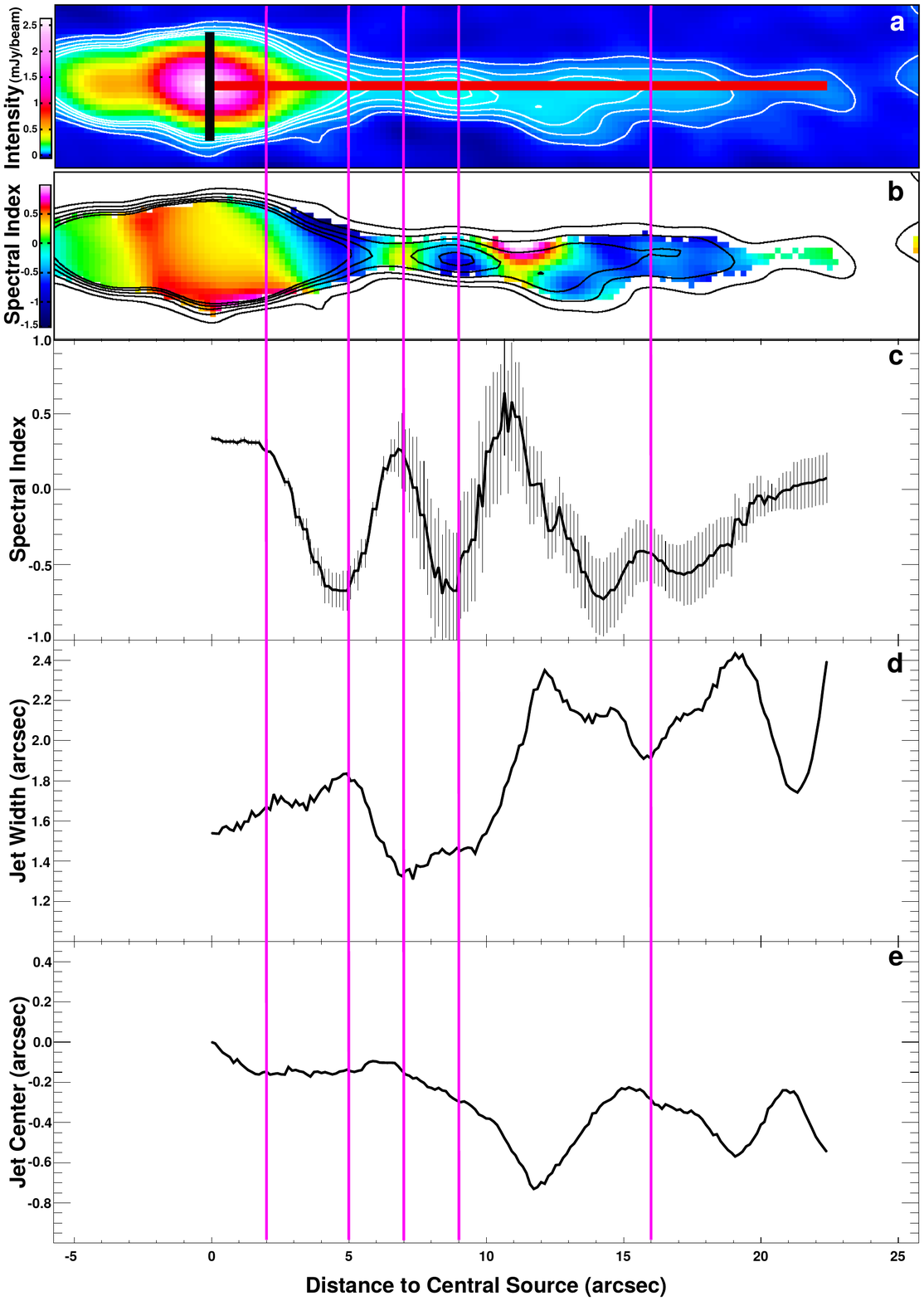}
  \caption{\footnotesize{Variation of jet parameters with distance z from the phase center along the collimated region. (a) Superimposed to the radio continuum image (L+S+C bands), two reference slices are shown. The red slice indicates the total range considered to compute parameter variations ($\sim$23 arcsec); the black slice represents a 3~arcsec cross-section for computing intensity and spectral index values along the collimated jet. Contours are 3, 5, 7, 9, 20 times the rms noise (9 $\mu$Jy/beam). (b) Superposition of the radio continuum image of panel {\it a} (contours) over the spectral index image obtained from multifrequency synthesis cleaning (color scale) and robust weighting 1. Panels (c), (d), and (e) show the variation of spectral index, jet width, and jet central axis with distance z, respectively. The jet width was estimated as the FWHM of a Gaussian fit to the brightness profile, while the jet central axis variations were computed with respect to the red slice. Vertical lines indicate distances of 2, 5, 7, 9, and 16 arcsec from the protostar.}}
  \label{Fig5}
  \end{figure}

 Our new images revealed an interesting behaviour of the spectral index in the most collimated part of the radio jet. We noted several changes in the spectral index from positive to negative and positive again. This is the first time that images of a radio jet with high quality enough are available to detect these changes. Therefore, in this subsection we deepen the analysis of the most collimated region of the jet, by studying how the spectral index, the jet width, and the position of the jet central axis vary along the distance z from the protostar. For this, we analize the highest signal-to-noise ratio images, combining L, S, and C bands. We defined a slice along the jet, from the central region of the jet to the farthest point in the northeastern lobe ($\alpha$(J2000)=18$^h$19$^m$13.227$^s$, $\delta$(J2000)=$-$20$^{\circ}$46$'$48.271$''$), with a total length of $\sim$45 arcsec. Centered on each pixel of this line, we took a 3~arcsec cross-section for computing intensity and spectral index values. For each transversal section, we estimated the jet width as the FWHM of a Gaussian fit to the intensity profile. Values of the spectral indices and their errors were measured in the pixels where the Gaussian fit center is located, while variations in the position of the jet central axis were computed with respect to the slice. The northeastern lobe region is much more complex than the most collimated one; the brigthest points are not aligned with the collimated jet, and the brightness profiles significantly differ from Gaussian fits. For this reason, we restricted our analysis to the region up to $\sim$23 arcsec from the phase center of the observations. Figure \ref{Fig5} shows intensity and spectral index images of the studied region, and the variations of spectral index, jet width, and jet central axis, with distance z.

Figure \ref{Fig5} shows that initially the jet widens, and its emission presents a positive spectral index, consistent with thermal free-free emission. At a distance of $\sim$2~arcsec from the center, the jet continues widening and the spectral index starts to decrease, reaching its most negative value ($\sim -0.5$) at $\sim$5~arcsec, roughly matching with the jet width relative maximum. Then, the jet starts to get narrower while the spectral index increases, reaching again positive values at $\sim$7~arcsec. Between $\sim$7 and $\sim$9 arcsec the spectral index decreases to negative values at the point where the jet width increases. At a distance of $\sim$16~arcsec, a relative maximum in the index is observed once more, coincident with a decreasing jet width.

Except for the thermal central region of the jet (see subsection \ref{nature}), the spectral index seems to decrease in widening regions and increase when the jet becomes narrower. We speculate that pressure differences between the jet and the ambient medium might give rise to expansion and recollimation of the jet. In this case, increasing spectral indices in narrow regions could be due to: 1) an increment in the opacity, 2) a local increase in thermal emission, through conversion of kinetic into internal energy in shocks. We also observe that the jet is suffering of lateral displacements, consistent with a precessing jet as observed by Mart\'{\i} et al. (1993). These displacements might also produce an increase in the spectral index, e.g. in regions where the jet approaches our line of sight, becoming apparently thicker and therefore, enlarging the integration depth of the radiation (and also the opacity if we assume a constant density). It is worth noting that these behaviours resemble those observed in recollimation shocks in AGN jets (e.g, Mimica et al. 2009, Fromm et al. 2013, Perucho 2013). Supersonic flows that are not in pressure equilibrium with their environment generate such shocks naturally: 1) the overpressured jet expands in the ambient medium, 2) once the jet pressure in the outer (radial) regions becomes smaller than the external pressure, recollimation starts from the jet boundary towards the axis, but 3) being the flow supersonic, the information crosses the jet as a shock. This process is natural to any supersonic jet flow, including YSO jets, and may trigger the observed structure. The interaction between shells of gas propagating at different speeds can trigger internal shocks in a jet. However, this would not imply changes in the jet width as significant as those given by jet overpressure (Fromm et al. 2016). Numerical simulations show that travelling shocks, if strong enough, can trigger local pressure enhancements and an increase of the jet cross section, which could be the source of the features known as trailing components (Agudo et al. 2001). However, there are a number of differences between travelling shocks (and trailing components) and standing, recollimation shocks: 1) the brightest spot would coincide with the presence of the travelling shock, i.e., with an expanded region, and not with the minimum in the jet radius (as observed in Fig. \ref{Fig5}), 2) the trailing features show a shorter wavelength than the distance between recollimation shocks, and 3) the travelling shocks and the trailing features propagate downstream. The observed evolution of the jet radius with distance favors in this case, which implies strong changes in the jet radius, our interpretation in terms of recollimation shocks. Nevertheless, in order to get a better understanding of the emission nature and structure of this highly collimated jet, new observations with higher angular resolution and higher sensitivity in small frequency intervals are needed.

\subsection{Other radio continuum sources}\label{other}

In addition to the central radio continuum source and the inner jet, we detected four new point-like sources in the field (Fig. \ref{Fig1}). The positions of these sources were measured in the highest resolution image (2.09$''\times$1.26$''$ at C band), and whenever possible, fluxes at different frequencies were measured to compute their spectral indices. The main parameters of these radio sources are given in Table \ref{tbl-NEW}. Source RC-1 presents $\alpha_{(18.4-10)\rm cm}>0$ and $\alpha_{(10-5.5)\rm cm}<0$, suggesting it could be the case of an HII region (emitting free-free radiation), optically thick at $\lambda>10$~cm and optically thin at $\lambda<10$~cm. For source RC-4 we obtain $\alpha_{(10-5.5)\rm cm}<0$ suggesting non-thermal emission, although within errors, optically thin free-free radiation is also possible. Sources RC-2 and RC-3 were reliably detected only at C band.

\section{On Particle Acceleration}\label{acc}

\begin{figure}
  \epsscale{1}
  \plotone{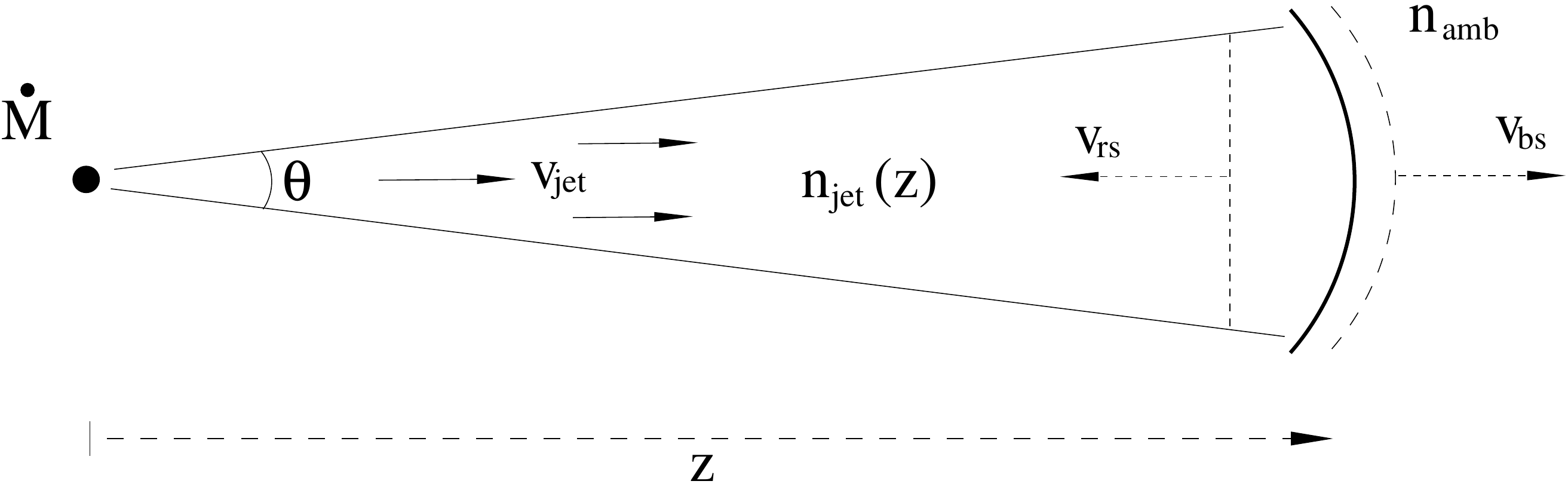}
  \caption{\footnotesize{Scheme showing parameters involved in the interaction of the jet with the ambient cloud. Adapted from Rodr\'iguez-Kamenetzky et al. (2016).}}
  \label{Fig6}
  \end{figure}

 Carrasco-Gonz\'alez et al. (2010) detected linearly polarized emission at 6 cm from the radio jet in HH 80-81. This confirmed the synchrotron nature of the radiation, implying the presence of relativistic particles. Theoretical models of DSA in jets from massive YSOs (Araudo et al. 2007, Bosch-Ramon et al. 2010) show that, under certain conditions, these sources can accelerate particles up to relativistic energies in strong adiabatic shocks produced by the interaction of the jets with the surrounding molecular cloud, as it is sketched in Figure \ref{Fig6}; (See also Padovani et al. 2015, 2016 for the case of low mass protostars). Based on the study of the triple source in Serpens (Rodr\'iguez-Kamenetzky et al. 2016) and HH 80-81 (this article), we infer that adiabatic reverse shocks can occur when the jet-to-ambient density ratio is smaller than unity ($n_{\rm jet}<n_{\rm amb}$), and jet velocities are moderately high compared with typical values (a few hundreds kilometers per second). Nevertheless, the mechanism to account for particle acceleration still remains unknown.
 
 By considering the simplest scenario where knot 1 indicates the jet termination region (see Figure \ref{Fig6}), we study whether electrons can be efficiently accelerated in both the bow shock and the reverse shock (Mach disk). In order to discern whether a shock is adiabatic or radiative we use the criterion adopted by Blondin et al. (1989). These authors define the dimensionless cooling parameter $\chi_{\rm s}\equiv d_{\rm cool}/r_{\rm jet}$ to compare the thermal cooling distance ($d_{\rm cool}$) with the jet radius (r$_{\rm jet}$) at the position of the shock. Here, $d_{\rm cool}$ represents the distance behind a steady state radiative shock to the point where the gas (that enters the shock at a velocity $v_{\rm s}$) has cooled to $\sim 10^{4}$~K. When the shock-heated gas does not have time to cool before leaving the working surface, $\chi_{\rm s}\gg 1$ and the shock is effectively adiabatic, whereas for a fully radiative shock, the postshock gas loses its thermal energy in a relatively short distance downstream of the shock, resulting $\chi_{\rm s}\ll 1$. For the jet radius we measured a value $r_{\rm jet}=1.21''=3.08\times 10^{16}$~cm corresponding to the semi minor axis of a Gaussian fit to the knot 1 in the wide bandwidth image (combining L, S, and C bands), see Fig. \ref{Fig4}. The cooling distance downstream of a shock with velocity $v_{\rm s}$ propagating in a medium with density $n$ can be estimated as:

\begin{equation}\label{cd1}
 \left(\frac{d_{\rm cool}}{\rm cm}\right)=1.8\times 10^{14}\left(\frac{100~{\rm cm}^{-3}}{n}\right)\left(\frac{v_{\rm s}}{100~\rm km~s^{-1}}\right)^{4.67};\hspace{10pt} v_{\rm s}>60~{\rm km~s}^{-1}
\end{equation}

\noindent (Hartigan et al. 1987) and

\begin{equation}\label{cd2}
 \left(\frac{d_{\rm cool}}{\rm cm}\right)=2.24\times 10^{14}\left(\frac{100~{\rm cm}^{-3}}{n}\right)\left(\frac{v_{\rm s}}{100~\rm km~s^{-1}}\right)^{4.5};\hspace{10pt} v_{\rm s}>400~{\rm km~s}^{-1}
\end{equation}

\noindent (Raga et al. 2002). In the following, we estimate the cooling distance in the bow shock and the Mach disk of knot 1.
  
\subsection{Bow Shock}
 
No proper motions for the identified knots are available at present, since no previous observations were able to resolve the structure of the radio jet. Therefore, the bow shock propagation velocity ($v_{\rm bs}$) needs to be indirectly estimated. Following Blondin et al. (1990), $v_{\rm bs}$ can be approximated by equating the momentum flux of the beam (the collimated supersonic flow) at the working surface with the momentum flux of the bow shock:

\begin{equation}\label{blondin}
v_{\rm bs}\approx\frac{v_{\rm jet}}{1+\eta^{-1/2}},\hspace{10pt}\eta\equiv\frac{n_{\rm jet}}{n_{\rm amb}}
\end{equation}

For the ambient density we adopt the average value measured by Torrelles et al. (1986): $n_{\rm amb}=5\times 10^{3}~{\rm cm}^{-3}$, and for the jet velocity we set $v_{\rm jet}=1000~{\rm km~s}^{-1}$ (Mart\'i et al. 1995). For the jet density at a distance {\it z} from the protostar we estimated a value $n(z_{\rm knot})=40$~cm$^{-3}$ (see Appendix \ref{appendix}). Therefore, from Eq. (\ref{blondin}) we find a bow-shock velocity $v_{\rm bs}=80~{\rm km~s}^{-1}$. According to this value, we calculate the cooling distance from Eq. (\ref{cd1}). This gives a $d_{\rm cool}/r_{\rm jet}$ ratio smaller than unity (4$\times$10$^{-5}$), which implies that the bow shock is radiative. In a radiative shock, Alfven waves are mostly suppressed and DSA is inefficient.

\subsection{Reverse Shock}

To study whether particle acceleration can be efficient in the reverse shock, we first estimate its velocity $v_{\rm rs}$. From conservation arguments in fluid equations, and assuming pressure equilibrium in shocked regions, the reverse shock velocity can be expresed as:

\begin{equation}\label{vrs}
v_{\rm rs}=v_{\rm jet}-3v_{\rm bs}/4
\end{equation}

\noindent Considering the values adopted in the previous section for $v_{\rm bs}$ and $v_{\rm jet}$, we found the reverse shock velocity to be $v_{\rm rs}\sim 940~{\rm km~s}^{-1}$. Therefore, the thermal cooling distance estimated from equation (\ref{cd2}) with $n=n_{\rm jet}$ gives $d_{\rm cool}/r_{\rm jet}=432$, implying an adiabatic reverse shock, that, in principle, might produce efficient particle acceleration up to relativistic energies. 


\subsection{Particle acceleration efficiency and energetics}

 We shall now explicitly show that the source can produce the relativistic electrons required to generate the observed synchrotron radiation and we shall discuss the efficiency of the process.

 The acceleration time scale for electrons in the reverse shock is:
 
 \begin{equation}
  t_{\rm acc}=\eta\frac{E}{eBc},
 \end{equation}

 \begin{equation}
  {\rm with} \hspace{10pt}\eta\sim 20\frac{D}{r_{\rm g}c}\left(\frac{c}{v_{\rm rs}}\right)^{2}.
 \end{equation}
 
 In these expressions $B$ is the magnetic field strength, $E$ is the energy of the particles, $v_{\rm rs}$ is given by eq. \ref{vrs} above, and $D$ is the diffusion coefficient. In the Bohm limit $D\sim D_{\rm B}=r_{\rm g}\frac{c}{3}$, with $r_{\rm g}$ the gyro-radius of the particles being accelerated. Thus,
 
  \begin{equation}
   \eta\sim\frac{20}{3}\left(\frac{c}{v_{\rm rs}}\right)^{2}.
  \end{equation}

 If the synchrotron losses dominate, the maximum energy of electrons -$E_{\rm max}$- is determined by the condition $t_{\rm acc}=t_{\rm synch}$, where $t_{\rm synch}$ is the time scale of synchrotron losses. In suitable units:
 
 \begin{equation}
  t_{\rm synch}\sim 4\times10^{11}\left(\frac{B}{\rm mG}\right)^{-2}\left(\frac{E}{\rm GeV}\right)^{-1}\hspace{10pt}\rm{s}.
 \end{equation}
 
  Thus, 
 
 \begin{equation}
 E_{\rm max}\sim 2.4\times10^{3}\left(\frac{v_{\rm rs}}{10^{8}\rm{cm~s}^{-1}}\right)\left(\frac{B}{\rm mG}\right)^{-\frac{1}{2}}\hspace{10pt}\rm{GeV}.
 \end{equation}

 The strength of the magnetic field and the diffusion regime are unknown. We consider a magnetic field strength of the order of 1~mG (Carrasco-Gonz\'alez et al. 2010), and Bohm diffusion. Thus, the maximum energy of the electrons results in $E_{\rm max}\sim2$~TeV. This means Lorentz factors of $\gamma\sim4\times10^{6}$ for electrons. This value should be considered as an upper limit, since other losses, such as relativistic Bremsstrahlung are present.  In any case, the energies necessary to produce the observed nonthermal radio emission by synchrotron mechanism (of the order of $\sim$GeV) are easily achieved in this source. This seems not to be the case in other YSOs with slower jets and denser media. In particular, if the particle density of not-fully ionized atoms is large, ionization losses can quench the acceleration (Bosch-Ramon et al. 2010, Padovani et al. 2015, Padovani et al. 2016). The crucial point is that if the shock is radiative, instead of adiabatic, a lot of heat will be generated in the shocked region and this will dramatically increase the entropy. This means that all order will be destroyed, in particular the inhomogeneities in the magnetic field that make the diffusive particle acceleration possible. Since the condition of adiabaticity is that the thermal time scale $t_{\rm therm}\propto n^{-2}$ be longer than the shock time scale $t_{\rm sh}$ (or equivalently $d_{\rm cool}>r_{\rm jet}$), the diffusive shock acceleration will be halted by either a slow shock or a dense medium. In this sense, the conditions in HH~80-81 seem to be quite special, making of it an almost unique laboratory for particle acceleration in YSO jets. 

 The synchrotron luminosity that we can expect from the electrons will depend on the efficiency for transforming shock power to non-thermal particles and the fraction that goes to electrons. Currently, it is not clear how efficient particle acceleration is in the case of shocks in protostellar jets. However, we can obtain some information about cosmic rays received at the Earth and compare with the better understood case of shocks in Galactic supernovae in order to arrive at some preliminary estimates.
  
 The energy density of Galactic cosmic rays (CR) around the Earth is explained if 1\% -- 10\% of the explosion energy of the Galactic supernovae is used for CR acceleration (Ginzburg \& Syrovatskii 1964). Detailed studies of particular supernova remnants such as those of the northeast region of the young SNR RCW 86  (Helder et al. 2009) suggest extremely efficient CR acceleration with $\eta\approx87$ \%, for shocks with velocities of $\sim6000$ km s$^{-1}$. Even if the shock velocity is as low as 1200 km s$^{−1}$ the efficiency would be 18\%. Recent investigations using three-dimensional magnetohydrodynamics simulations of supernova shock waves propagating into a realistic, diffuse medium by Shimoda et al. (2015) show that the above values might be overestimated by 10\% -- 40\% because a rippled shock does not immediately dissipate all of the upstream kinetic energy. On this basis, efficiencies of $\sim 10$ \% for shocks of $\sim 1000$ km~s$^{-1}$ moving through a complex medium as it is the case of HH~80-81 jet seem to be reasonable, i.e. in accord with both observational inferences in a variety of circumstances and theoretical expectations (see, for instance, Bell 2013, Blasi 2013, Morlino et al. 2013, 2014). 
 
 The composition of cosmic rays in the solar neighbourhood indicates that about 1\% of them are electrons. The luminosity ratio of electrons to protons as it is produced in stochastic acceleration processes in different sources is an important quantity relevant for several aspects of the modeling of the sources. The mentioned ratio of 1\% is usually assumed to be valid for the case of Galactic sources.  Different types of observations show that the average ratios should be close to this value (although fluctuations will occur depending on specific conditions of the source, see Merten et al. 2017).

 In order to obtain an estimation of the expected synchrotron luminosity in HH 80-81, we first derive the synchrotron bolometric luminosity of the shock, following Bosch-Ramon et al. (2010):
 
 \begin{equation}
  \frac{L_{\rm shock}}{{\rm erg~s}^{-1}}\sim10^{35}\left(\frac{R}{3\times10^{16}~{\rm cm}}\right)^{2}\left(\frac{v_{\rm rs}}{1000~{\rm km~s}^{-1}}\right)^{3}\left(\frac{n_{\rm j}}{40~{\rm cm}^{-3}}\right)
 \end{equation}

 \noindent In accordance with the above considerations we adopt an efficiency of 10 \% for the particle acceleration in the shocks of HH~80-81, and we assume that $\sim 1 $ \% of them are electrons. Thus, we get a bolometric luminosity $L_{\rm synch}\sim10^{32}\hspace{10pt}\rm{erg~s}^{-1}$, which is well above the vaule inferred from the radio observations ($L_{\rm (1-6) GHz}\sim~9\times10^{27}{\rm erg~s}^{-1}$, synchrotron luminosity integrated in the 1-6 GHz spectral range). Therefore, the observed synchrotron emission seems to be easily explained with particle acceleration in strong shocks of the jet against the ambient medium assuming reasonable parameters.

\section{Conclusions}
  
 We have presented an analysis of new JVLA observations of the HH 80-81 inner jet. Our study of the jet morphology and the nature of its radio emission through wide bandwidth multifrequency data, lead us to the following conclusions:

 \begin{enumerate}

  \item The inner jet is detected at all the frequencies analyzed with a full extension of $\sim 1~{\rm pc}$. At low frequencies and low angular resolution, a very extended lobe-like morphology is observed to the northeast and southwest of the exciting source, while for increasing frequencies and higher angular resolution, a very well collimated jet and high intensity spots are revealed.

  \item In the most collimated region an alternated spatial distribution of positive and negative spectral indices are observed, while variations of the jet width are measured. The general trend seems to indicate that the jet reaches negative spectral indices in widening regions, and positive spectral indices when it gets narrow. We speculate that pressure differences between the jet and the ambient medium might give rise to expansion and recollimation of the jet; a behaviour that resembles what is observed in recollimation shocks in AGN jets.

  \item Spectral indices measured for the central region of the jet are consistent with partially optically thick free-free emission. At low frequencies the extended emission is characterized by negative spectral indices, suggesting the presence of non-thermal radiation. With increasing frequencies, positive and negative spectral indices are measured, indicating both thermal and non-thermal contributions. Nevertheless, high intensity spots (knot 1 and 2) in the northeastern lobe reveal their non-thermal origin ($\alpha=-0.6\pm0.2$) in the highest resolution image in the 2-4~GHz range, for which we obtain the lower rms.
   
  \item We discuss whether particle acceleration can be efficient in both the forward and the reverse shocks of the high intensity knot 1, and conclude that electrons could be accelerated via the DSA mechanism in the reverse shock.

 \end{enumerate}

 The work of A.R-K. and C.C-G. was supported by UNAM-DGAPA-PAPIIT grant number IA101214 and MINCyT-CONACyT ME/13/47 grant, corresponding to the Bilateral Cooperation Program. G.A. and J.M.T. acknowledge support from MINECO (Spain) AYA2014-57369-C3 grant (co-funded by FEDER). GER acknowledges support from by the Argentine Agency CONICET (PIP 2014-00338) and the Spanish Ministerio de Econom\'ia y Competitividad (MINECO/FEDER, UE) under grants AYA2013- 47447-C3-1-P and AYA2016-76012-C3-1-P. J.M. ackenowledges support from Spanish MINECO and Consejer\'{\i}a de Innovaci\'on, Ciencia y Empresa grants AYA2016-76012-C3-3-P and FQM-322 (co-funded by FEDER). MP acknowledges support by the Spanish ``Ministerio de Econom\'{\i}a y Competitividad'' grants AYA2013-40979-P, and AYA2013-48226-C3-2-P.

\newpage 
 \appendix
\section{Jet density in a bipolar stellar wind}\label{appendix}

\begin{figure}
  \epsscale{0.5}
  \plotone{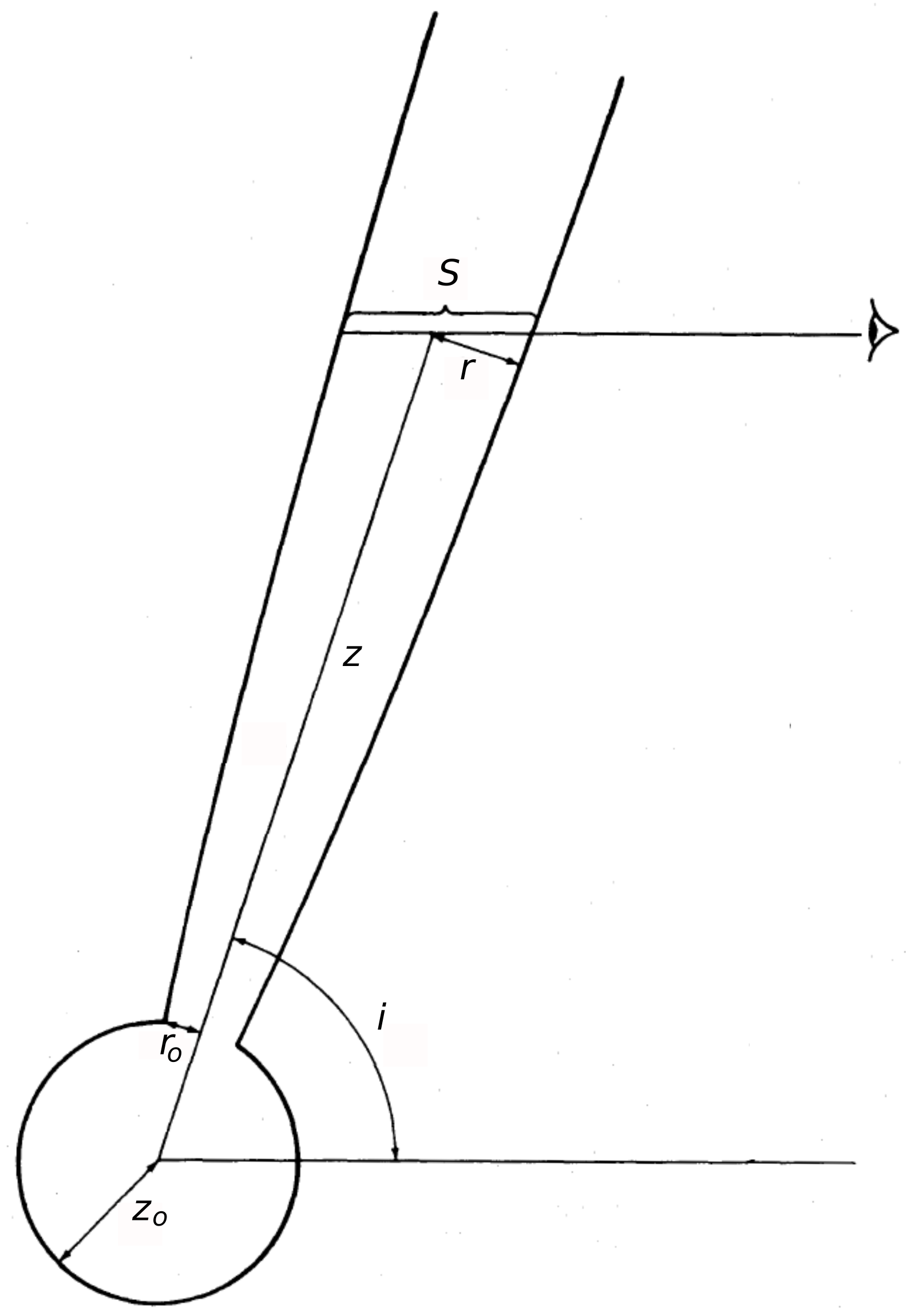}
  \caption{\footnotesize{Jet geometry. The flow is assumed to be injected at a distance $z_{o}$ where the jet radius is $r_{o}$. Figure adapted from Reynolds 1986.}}
  \label{Fig7}
  \end{figure}

  We consider a stellar wind (Fig. \ref{Fig7}), characterized by a mass loss rate ($\dot{M}$), ejection velocity ($v$), distance to the star where the ionized jet begins ($z_{0}$), and wind aperture semi-angle ($\theta_{0}$). From the conservation of mass in a fluid flow, the particle density ($n$) can be derived as a function of distance $z$ from the protostar. The continuity equation states:
 
  \begin{equation}\label{A1}
   \dot{M}=2S(z)~n~\mu~v
  \end{equation}
 
  \noindent being $S(z)$ the surface area crossed by the wind at a distance $z$ from the protostar, and $\mu$ the hydrogen mass. The factor 2 represents that the mass  lost by the protostar is distributed in two spherical sections. From equation \ref{A1} we can obtain an expression for the jet density at a distance $z$ from the driving source ($n=n(z)$), that in suitable units results:
 
  \begin{equation}\label{A2}
   \frac{n(z)}{\rm cm^{-3}}=3.15\times10^{20}\left(\frac{\dot{M}}{\rm M_{\odot}~yr^{-1}}\right)\left(\frac{\mu}{\rm gr}\right)^{-1}\left(\frac{v}{\rm km~s^{-1}}\right)^{-1}\left(\frac{S(z)}{\rm cm^{2}}\right)^{-1}
  \end{equation}

  In order to estimate $n(z)$, we first need to calculate $\dot{M}$. With this purpose we follow Eq. 3 from Beltr\'an et al. (2001), based on the article of Reynolds (1986):
 
   \begin{equation}\label{A3}
   \begin{aligned}
   \left(\frac{\dot{M}}{10^{-6}~\rm M_{\odot}~yr^{-1}}\right)=0.108\left[\frac{(2-\alpha)(0.1+\alpha)}{1.3-\alpha}\right]^{0.75}\left[\left(\frac{S_{\nu}}{\rm mJy}\right)\left(\frac{\nu}{10~\rm GHz}\right)^{-\alpha}\right]^{0.75}\\
   \times\left(\frac{V_{*}}{200~\rm km~ s^{-1}}\right)\left(\frac{\nu_{m}}{10~\rm GHz}\right)^{0.75\alpha-0.45}\left(\frac{\theta_{o}}{\rm rad}\right)^{0.75}(\sin i)^{-0.25}\left(\frac{d}{\rm kpc}\right)^{1.5}\left(\frac{T}{10^{4}~\rm K}\right)^{-0.075},
  \end{aligned}
  \end{equation}

  \noindent Where $V_{*}$ is the wind velocity at the base of the jet, $\theta_{o}$ the jet injection opening angle, $i$ the jet axis inclination respect to the line of sight (assumed $\sim 90^{\circ}$), $T$ the temperature, $S_{\nu}$ the flux density of the source at the frequency $\nu$, below the turn-over frequency (above which the entire jet becomes optically thin), and $\alpha$ the spectral index of the source. All parameters involved in equations are listed in Table \ref{tbl-app}.

  The jet injection opening angle $\theta_{o}$ is related with the injection jet half-width $r_{o}$, at a distance $z_{o}$ (where the ionized gas density is highest): $\tan\theta_{o}=r_{o}/z_{o}$. Both $r_{o}$ and $z_{o}$ are unknown, but, following Reynolds (1986), we assume a power law dependence of the jet half-width on jet length $z$:

  \begin{equation}\label{A4}
    r(z)=r_{o}\left(\frac{z}{z_{o}}\right)^{\epsilon}
  \end{equation}
 
  \noindent were $\epsilon$ is a parameter determined by the spectral index of the source ($\alpha=1.3-0.7\epsilon$, considering an isothermal, constant-velocity, and fully ionized flow, Reynolds 1986). By measuring the jet half-width $r_{\rm knot}$ at the position of knot 1 ($z_{\rm knot}$), and using Equation \ref{A4}, we can write an expression for $\theta_{o}$:
 
  \begin{equation}\label{A5}
   \theta_{o}=\tan^{-1}\left(\frac{r_{\rm knot}}{z_{\rm knot}^{\epsilon}}z_{o}^{\epsilon-1}\right)
  \end{equation}

  \noindent Therefore, assuming 10~au as initial value for $z_{o}$ in Eq. \ref{A5}, we obtain the first approximated value for $\theta_{o}=0.33$~rad. Thus, using the expression given by Beltr\'an et. al (2001) for $z_{o}$,
 
  \begin{equation}
  \begin{aligned}
    \left(\frac{z_{o}}{\rm au}\right)=26\left[\frac{(2-\alpha)(0.1+\alpha)}{1.3-\alpha}\right]^{0.5}\times\left[\left(\frac{S_{\nu}}{\rm mJy}\right)\left(\frac{\nu}{10~\rm GHz}\right)^{-\alpha}\right]^{0.5}\left(\frac{\nu_{m}}{10~\rm GHz}\right)^{0.5\alpha-1}\\
    \times\left(\frac{\theta_{o}\sin i}{\rm rad}\right)^{-0.5}\left(\frac{d}{\rm kpc}\right)\left(\frac{T}{10^{4}~\rm K}\right)^{-0.5}
    \end{aligned}
   \end{equation}

  \noindent we perform an iterative process to find the convergent values of $\theta_{o}$ and $z_{o}$, obtaining $\theta_{o}=0.2772$~rad~$\simeq16^{\circ}$ and $z_{o}=20$~au. We also tried different initial conditions for $z_{o}$ and the result was always the same, indicating stable convergence. Then, from Equation \ref{A3} we obtain a mass loss rate $\dot{M}=6\times10^{-7}~{\rm M_{\odot}~yr^{-1}}$, and finally, with Equation \ref{A2} we calculate the jet density at the position of knot 1 $n(z_{\rm knot})=40$~cm$^{-3}$.


\newpage


  \begin{deluxetable}{ccccccc}
  \tabletypesize{\normalsize}
  \tablecaption{PARAMETERS OF THE IMAGES \label{tbl-images}}
  \tablewidth{0pt}
  \tablehead{
  \colhead{\centering Epoch} & 
  \colhead{Spectral} &  
  \colhead{Configuration} &
  \colhead{Bandwidth} &  
  \colhead{Weighting} & 
  \colhead{Synthesized} & 
  \colhead{PA} \\
  \colhead{} &
  \colhead{Band} &
  \colhead{ } &
  \colhead{(GHz)} &
  \colhead{ } &
  \colhead{Beam} &
  \colhead{ }
  }
  \startdata
  2012$^{a}$              &    L          &    B    &    1.0      &    Robust = 1      &    $7.17''\times 4.30''$   &   7$^\circ$        \\   
  2012$^{a}$              &    S          &    B    &    2.0      &    Robust = 1      &    $4.10''\times 2.21''$   &   -1$^\circ$       \\      
  2012$^{a}$              &    C          &    B    &    2.0      &    Robust = 1      &    $2.09''\times 1.26''$   &   5$^\circ$        \\
  2012$^{a}$              &    L+S        &    B    &    3.0      &    Robust = 1      &    $4.81''\times 2.73''$   &   0$^\circ$     \\      
  2012$^{a}$              &    S+C        &    B    &    4.0      &    Robust = 1      &    $2.56''\times 1.37''$   &   5$^\circ$        \\
  2012$^{a}$              &    S+C        &    B    &    4.0      &    Robust = 0      &    $1.77''\times 1.03''$   &   4$^\circ$     \\
  2012$^{a}$              &    L+S+C      &    B    &    5.0      &    Robust = 1      &    $2.37''\times 1.46''$   &   4$^\circ$     \\
  2009$^{b}$              &    C    &          D    &    0.1       &   Natural    
       &    13$''$ $\times$ 8$''$   &    2$^\circ$
  \enddata
  \tablenotetext{a}{Combined data from June 12 and 16, 2012.}
  \tablenotetext{b}{From Carrasco-Gonz\'alez et al. 2010.}

 \end{deluxetable}   

  \begin{deluxetable}{cccccccc}
  \tabletypesize{\footnotesize}
  \tablecaption{PARAMETERS OF NEW SOURCES\label{tbl-NEW}}
  \tablewidth{0pt}
  \tablehead{
  \colhead{} & 
  \multicolumn2c{Position} &
  \multicolumn3c{Flux ($\mu$Jy)}&
  \multicolumn2c{Spectral Index}\\
  \cline{2-8}
  \colhead{\centering SOURCE} & 
  \colhead{$\alpha$(J2000)} &  
  \colhead{$\delta$(J2000)} &
  \colhead{18.4 cm} &  
  \colhead{10 cm} & 
  \colhead{5.5 cm} & 
  \colhead{$\alpha_{(18.4-10)\rm cm}$} &
  \colhead{$\alpha_{(10-5.5)\rm cm}$}
  }
   \startdata
  RC-1$^a$ ...........  &  18$^{h}$19$^{m}$10.99$^{s}$  &  $-$20$^{\circ}$48$^{\prime}$29.59$^{\prime\prime}$  &  444 $\pm$ 56  &  662 $\pm$ 78  &  481 $\pm$ 48  &  $+$0.7 $\pm$ 0.3 &  $-$0.5 $\pm$ 0.2 \\
  RC-2$^b$ ...........  &  18 19 12.48  &  $-$20 47 27.30  &  --  &  --  &  61 $\pm$ 8  &  --  &  --  \\
  RC-3$^b$ ...........  &  18 19 12.58  &  $-$20 47 30.45  &  --  &  --  &  100 $\pm$ 8  &  --  &  --  \\
  RC-4$^c$ ...........  &  18 19 12.93  &  $-$20 47 37.77  &  --  &  81 $\pm$ 6  &  60 $\pm$ 8  &  -- & $-$0.5 $\pm$ $0.3$
   \enddata
  \tablenotetext{a}{Flux densities are obtained from Gaussian fits to L, S, and C images convolved to the same beam size (7.3$''\times$ 4.4$''$, PA$=$7$^\circ$).}
  \tablenotetext{b}{Flux densities are obtained from Gaussian fits to C band image.}
  \tablenotetext{c}{Flux densities are obtained from Gaussian fits to S, and C images convolved to the same beam size (4.2$''\times$ 2.3$''$, PA$=$0$^\circ$).}

  \tablecomments{All positions are measured from a Gaussian fit to C band image, with the highest resolution (2.09$''\times$ 1.26$''$, PA$=$5$^\circ$).}
 \end{deluxetable}

 
   \begin{deluxetable}{lll}
  \tabletypesize{\small}
  \tablecaption{PARAMETERS INVOLVED IN EQUATIONS \label{tbl-app}}
  \tablewidth{0pt}
  \tablehead{
  \colhead{Parameter} & 
  \colhead{Value} &  
  \colhead{Description}
  }
  \startdata
  $\alpha\hspace{3pt}^{\rm a}$         &    0.33                                               &   {\small Spectral index} \\
  $\epsilon$                           &    0.72                                               &   {\small Jet half-width power law index} \\
  $S_{\nu}\hspace{3pt}^{\rm a}$        &    2.63~mJy                                           &   {\small Flux density} \\
  $\nu$                                &    3.880287~GHz                                       &   {\small Reference fequency} \\
  $\nu_{\rm m}$                        &    100 GHz                                            &   {\small Turn-over frequency} \\
  $\theta_{o}$                         &    0.2772~rad                                         &   {\small Jet injection opening angle} \\
  $z_{o}$                              &    20~au                                              &   {\small Jet injection distance} \\
  T                                    &    $10^{4}$~K                                         &   {\small Electron temperature} \\
  d                                    &    1.7~kpc                                            &   {\small Distance to the HH 80-81 system} \\
  $\mu$                                &    1.6735$\times10^{-24}$~gr                          &   {\small Hydrogen mass} \\
  S(z)\hspace{3pt}$^{\rm b}$           &    2.8$\times10^{33}$~cm$^2$                          &   {\small Jet cross section} \\
  $w_{\rm knot}\hspace{3pt}^{\rm c}$   &    2057~au=$3\times10^{16}$~cm=$1.21''$               &   {\small Jet half-with} \\
  $z_{\rm knot}\hspace{3pt}^{\rm d}$   &    71400~au=$1\times10^{18}$~cm=$z=42''\sim0.35$~pc   &   {\small Distance of knot 1 to the driving source}   
  
  \enddata
  \tablenotetext{a}{\small The values of $\alpha$, $S_{\nu}$, and $\nu$, correspond to the L+S+C image.}
  \tablenotetext{b}{\small Transversal section crossed by the jet at $z_{\rm knot}$.}
  \tablenotetext{c}{\small Semi minor axis of a Gaussian fit to knot 1 in the L+S+C image.}
  \tablenotetext{d}{\small Measured in the L+S+C image.}

 \end{deluxetable}
 
 
\end{document}